\documentclass{article}
\usepackage{graphicx,amsfonts,amsmath,amssymb,hyperref}
\input{tcilatex}
\begin{document}

\title{Human--Robot Biodynamics}
\author{Vladimir G. Ivancevic \\
Land Operations Division, Defence Science \& Technology
Organisation\\ P.O. Box 1500, Edinburgh SA 5111, Australia}
\date{}
\maketitle

\begin{abstract}
This paper presents the scientific body of knowledge behind the
Human Biodynamics Engine (HBE), a human motion simulator developed
on the concept of Euclidean motion group SE(3), with 270 active
degrees of freedom, force--velocity--time muscular mechanics and
two--level neural control -- formulated in the fashion of
nonlinear humanoid robotics. The following aspects of the HBE
development are described: geometrical, dynamical, control,
physiological, AI, behavioral and
complexity, together with several simulation examples.\\

biodynamics,
Lie-derivative control, muscular mechanics, fuzzy--topological
coordination,\\ biodynamical complexity, validation, application
\end{abstract}

\section{Introduction}

Both human biodynamics and humanoid robotics are devoted to studying
human--like motion. They are both governed by Newtonian dynamical laws and
reflex--like nonlinear controls \cite%
{GaneshIEEE,IJMMS1,SIAM,IJMMS2,LieLagr,GaneshSprSml,GaneshWSc,GaneshSprBig}.

Although, current humanoid robots more and more resemble human
motion, we still need to emphasize that human joints are (and will
probably always remain) significantly more flexible than humanoid
robot joints. Namely, each humanoid joint consists of a pair of
coupled segments with only Eulerian rotational degrees of freedom.
On the other hand, in each human synovial joint, besides gross
Eulerian rotational movements (roll, pitch and yaw), we also have
some hidden and restricted translations along $(X,Y,Z)-$axes. For
example, in the knee joint, patella (knee cap) moves for about
7--10 cm from maximal extension to maximal flexion). It is
well--known that even greater are translational amplitudes in the
shoulder joint. In other words, within the realm of rigid body
mechanics, a segment of a human arm or leg is not properly
represented as a rigid body fixed at a certain point, but rather
as a rigid body hanging on rope--like ligaments. More generally,
the whole skeleton mechanically represents a system of flexibly
coupled rigid bodies, technically an anthropomorphic topological
product of SE(3)--groups. This implies the more complex
kinematics, dynamics and control than in the case of humanoid
robots \cite{VladSanjeev}.

This paper presents the scientific body of knowledge behind the
sophisticated human motion simulator, formulated in the fashion of
nonlinear humanoid robotics, called the Human Biodynamics Engine
(HBE), designed over the last five years by the present author at
Defence Science \& Technology Organisation, Australia. The HBE is
a sophisticated human neuro-musculo-skeletal dynamics simulator,
based on generalized Lagrangian and Hamiltonian mechanics and
Lie-derivative nonlinear control. It includes 270 active degrees
of freedom (DOF), without fingers: 135 rotational DOF are
considered active, and 135 translational DOF are considered
passive. The HBE incorporates both forward and inverse dynamics,
as well as two neural--like control levels. Active rotational
joint dynamics is driven by 270 nonlinear muscular actuators, each
with its own excitation--contraction dynamics (following
traditional Hill--Hatze biomechanical models). Passive
translational joint dynamics models visco-elastic properties of
inter-vertebral discs, joint tendons and muscular ligaments as a
nonlinear spring-damper system. The lower neural control level
resembles spinal--reflex positive and negative force feedbacks,
resembling stretch and Golgi reflexes, respectively. The higher
neural control level mimics cerebellum postural stabilization and
velocity target-tracking control. The HBE's core is the full spine
simulator, considering human spine as a chain of 26
flexibly--coupled rigid bodies (formally, the product of 26
SE(3)--groups). The HBE includes over 3000 body parameters, all
derived from individual user data, using standard biomechanical
tables. The HBE incorporates a new theory of soft
neuro-musculo-skeletal injuries, based on the concept of the local
rotational and translational {jolts}, which are the time rates of
change of the total forces and torques localized in each joint at
a particular time instant.

\section{Geometrical Formalism of Human--Robot Biodynamics}

\subsection{Configuration Manifold of Idealistic Robot Motion}

\begin{figure}[tbh]
\centering \includegraphics[width=12cm]{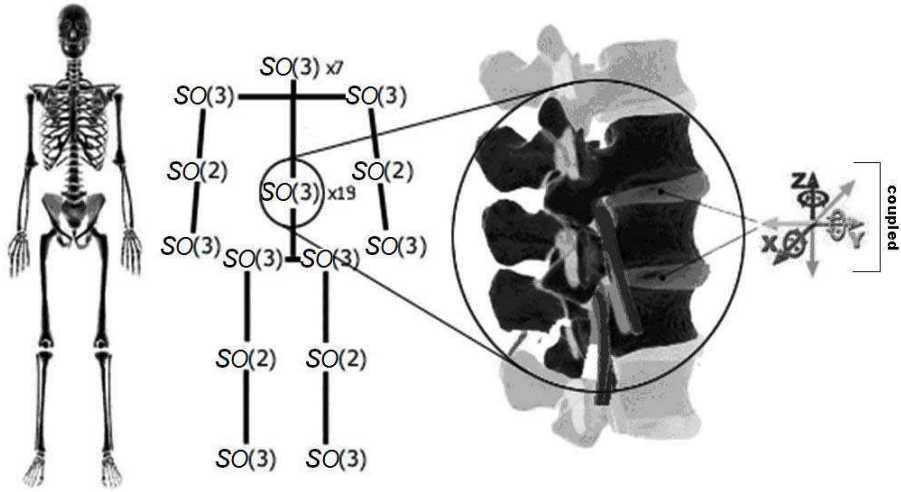} \caption{The
configuration manifold $Q_{rob}$ of the humanoid--robot body is
defined as a topological product of constrained $SO(3)$ groups, $%
Q_{rob}=\prod_{i}SO(3)^{i}$.}
\label{SpineSO(3)}
\end{figure}

Representation of an ideal humanoid--robot motion is rigorously defined in
terms of {rotational} constrained $SO(3)$--groups \cite%
{GaneshSprSml,GaneshSprBig,GaneshADG} in all main robot joints (see Figure %
\ref{SpineSO(3)}). Therefore, the {configuration manifold}
$Q_{rob}$
for humanoid dynamics is defined as a topological product of all included $%
SO(3)$ groups, $Q_{rob}=\prod_{i}SO(3)^{i}$. Consequently, the
natural stage for autonomous Lagrangian dynamics of robot motion
is the {tangent
bundle} $TQ_{rob}$\footnote{%
In mechanics, to each $n-$dimensional ($n$D) {configuration
manifold} $Q$ there is associated its $2n$D {velocity phase--space
manifold}, denoted by $TM$ and called the tangent bundle of $Q$.
The original smooth manifold $Q$ is called the {base} of $TM$.
There is an onto map $\pi :TM\rightarrow Q$, called the
{projection}. Above each point $x\in Q$ there is a {tangent space}
$T_{x}Q=\pi ^{-1}(x)$ to $Q$ at $x$, which
is called a {fibre}. The fibre $T_{x}Q\subset TM$ is the subset of $%
TM $, such that the total tangent bundle,
$TM=\dbigsqcup\limits_{m\in Q}T_{x}Q$, is a {disjoint union} of
tangent spaces $T_{x}Q$ to $Q$ for all points $x\in Q$. From
dynamical perspective, the most important quantity in the tangent
bundle concept is the smooth map $v:Q\rightarrow TM$, which
is an inverse to the projection $\pi $, i.e, $\pi \circ v=\func{Id}%
_{Q},\;\pi (v(x))=x$. It is called the {velocity vector--field}.
Its graph $(x,v(x))$ represents the {cross--section} of the
tangent
bundle $TM$. This explains the dynamical term {velocity phase--space}%
, given to the tangent bundle $TM$ of the manifold $Q$. The tangent bundle
is where tangent vectors live, and is itself a smooth manifold.
Vector--fields are cross-sections of the tangent bundle.
\par
System's {Lagrangian} (energy function) is a natural energy
function on the tangent bundle.} \cite{LieLagr}, and for the
corresponding autonomous
Hamiltonian dynamics is the {cotangent bundle} $T^{\ast }Q_{rob}$%
\footnote{%
A {dual} notion to the tangent space $T_{m}Q$ to a smooth manifold
$Q$ at a point $m$ is its {cotangent space} $T_{m}^{\ast }Q$ at
the same point $m$. Similarly to the tangent bundle, for a smooth
manifold $Q$ of dimension $n$, its {cotangent bundle} $T^{\ast }Q$
is the disjoint union of all its cotangent spaces $T_{m}^{\ast }Q$
at all points $m\in Q$, i.e., $T^{\ast }Q=\dbigsqcup\limits_{m\in
Q}T_{m}^{\ast }Q$. Therefore, the cotangent bundle of an
$n-$manifold $Q$ is the vector bundle $T^{\ast }Q=(TM)^{\ast }$,
the (real) dual of the tangent bundle $TM$. The cotangent bundle
is where 1--forms live, and is itself a smooth manifold.
Covector--fields (1--forms) are cross-sections of the cotangent
bundle. The {Hamiltonian} is a natural energy function on the
tangent bundle.} \cite{IJMMS1,SIAM}.

More precisely, the three--axial $SO(3)-$group of humanoid--robot joint
rotations depends on three parameters, Euler joint angles $q^{i}=(\varphi
,\psi ,\theta ),$\ defining the rotations about the Cartesian coordinate
triedar $(x,y,z)$ placed at the joint pivot point. Each of the Euler angles
are defined in the constrained range $(-\pi ,\pi )$, so the joint group
space is a constrained sphere of radius $\pi $ \cite%
{GaneshSprSml,GaneshSprBig,GaneshADG}.

Let $G=SO(3)=\{A\in \mathcal{M}_{3\times 3}(\mathbb{R}):A^{t}A=I_{3},\det
(A)=1\}$ be the group of rotations in $\mathbb{R}^{3}$. It is a Lie group
and $\dim(G)=3$. Let us isolate its one--parameter joint subgroups, i.e.,
consider the three operators of the finite joint rotations $R_{\varphi
},R_{\psi },R_{\theta }\in SO(3),$ given by
\begin{equation*}
R_{\varphi } =\left[
\begin{array}{ccc}
1 & 0 & 0 \\
0 & \cos \varphi & -\sin \varphi \\
0 & \sin \varphi & \cos \varphi%
\end{array}
\right] , ~~ R_{\psi } =\left[
\begin{array}{ccc}
\cos \psi & 0 & \sin \psi \\
0 & 1 & 0 \\
-\sin \psi & 0 & \cos \psi%
\end{array}
\right] , ~~ R_{\theta } =\left[
\begin{array}{ccc}
\cos \theta & -\sin \theta & 0 \\
\sin \theta & \cos \theta & 0 \\
0 & 0 & 1%
\end{array}
\right]
\end{equation*}
corresponding respectively to rotations about $x-$axis by an angle $\varphi
, $ about $y-$axis by an angle $\psi ,$ and about $z-$axis by an angle $%
\theta $.

The total three--axial joint rotation $A$ is defined as the product of above
one--parameter rotations $R_{\varphi },R_{\psi },R_{\theta },$ i.e., $%
A=R_{\varphi }\cdot R_{\psi }\cdot R_{\theta }$ is equal\footnote{%
Note that this product is noncommutative, so it really depends on the order
of multiplications.}
\begin{equation*}
A=\left[
\begin{array}{ccc}
\cos \psi \cos \varphi -\cos \theta \sin \varphi \sin \psi & \cos \psi \cos
\varphi +\cos \theta \cos \varphi \sin \psi & \sin \theta \sin \psi \\
-\sin \psi \cos \varphi -\cos \theta \sin \varphi \sin \psi & -\sin \psi
\sin \varphi +\cos \theta \cos \varphi \cos \psi & \sin \theta \cos \psi \\
\sin \theta \sin \varphi & -\sin \theta \cos \varphi & \cos \theta%
\end{array}
\right] .
\end{equation*}
However, the order of these matrix products matters: different order
products give different results, as the matrix product is {%
noncommutative product}. This is the reason why Hamilton's {%
quaternions}\footnote{%
Recall that the set of Hamilton's {quaternions} $\mathbb{H}$
represents an extension of the set of complex numbers
$\mathbb{C}$. We can
compute a rotation about the unit vector, $\mathbf{u}$ by an angle $\theta $%
. The quaternion $q$ that computes this rotation is
\begin{equation*}
q=\left( \cos \frac{\theta }{2}~,~u\sin \frac{\theta }{2}\right) .
\end{equation*}%
} are today commonly used to parameterize the $SO(3)-$group, especially in
the field of 3D computer graphics.

The one--parameter rotations $R_{\varphi },R_{\psi },R_{\theta }$ define
curves in $SO(3)$ starting from $I_{3}={\small \left(
\begin{array}{ccc}
1 & 0 & 0 \\
0 & 1 & 0 \\
0 & 0 & 1%
\end{array}
\right)} .$ Their derivatives in $\varphi =0,\psi =0$ and $\theta =0\,\ $%
belong to the associated {tangent Lie algebra} $\mathfrak{so}(3)$.
That is the corresponding infinitesimal generators of joint
rotations -- joint angular velocities $v_{\varphi },v_{\psi
},v_{\theta }\in \mathfrak{so}(3)$ -- are respectively given by
\begin{eqnarray*}
v_{\varphi } &=&{\small \left[
\begin{array}{ccc}
0 & 0 & 0 \\
0 & 0 & -1 \\
0 & 1 & 0%
\end{array}%
\right]} =-y\frac{\partial }{\partial z}+z\frac{\partial }{\partial y}%
,\qquad v_{\psi }={\small \left[
\begin{array}{ccc}
0 & 0 & 1 \\
0 & 0 & 0 \\
-1 & 0 & 0%
\end{array}%
\right]} =-z\frac{\partial }{\partial x}+x\frac{\partial }{\partial z}, \\
v_{\theta } &=&{\small \left[
\begin{array}{ccc}
0 & -1 & 0 \\
1 & 1 & 0 \\
0 & 0 & 0%
\end{array}%
\right]} =-x\frac{\partial }{\partial y}+y\frac{\partial }{\partial x}.
\end{eqnarray*}
Moreover, the elements are linearly independent and so
\begin{equation*}
\mathfrak{so}(3)=\left\{ \left[
\begin{array}{ccc}
0 & -a & b \\
a & 0 & -\gamma \\
-b & \gamma & 0%
\end{array}
\right] |a,b,\gamma \in \mathbb{R}\right\}.
\end{equation*}
The Lie algebra $\mathfrak{so}(3)$ is identified with $\mathbb{R}^{3}$ by
associating to each $v=(v_{\varphi },v_{\psi },v_{\theta })\in \mathbb{R}%
^{3} $ the matrix $v\in \mathfrak{so}(3)$ given by $v={\small \left[
\begin{array}{ccc}
0 & -a & b \\
a & 0 & -\gamma \\
-b & \gamma & 0%
\end{array}
\right]} . $ Then we have the following identities:

\begin{enumerate}
\item $\widehat{u\times v}=[\hat{u},v]$; ~and

\item $u\cdot v=-\frac{1}{2}\limfunc{Tr}(\hat{u}\cdot v)$.
\end{enumerate}

The exponential map $\exp :\mathfrak{so}(3)\rightarrow SO(3)$ is
given by {Rodrigues relation}
\begin{equation*}
\exp (v)=I+\frac{\sin \left\Vert v\right\Vert }{\left\Vert v\right\Vert }v+%
\frac{1}{2}\left( \frac{\sin \frac{\left\Vert v\right\Vert }{2}}{\frac{%
\left\Vert v\right\Vert }{2}}\right) ^{2}v^{2},
\end{equation*}
where the norm $\left\Vert v\right\Vert $ is given by
\begin{equation*}
\left\Vert v\right\Vert =\sqrt{(v^{1})^{2}+(v^{2})^{2}+(v^{3})^{2}}.
\end{equation*}

The dual, {cotangent Lie algebra} $\mathfrak{so}(3)^{\ast },$
includes the three joint angular momenta $p_{\varphi },p_{\psi
},p_{\theta }\in \mathfrak{so}(3)^{\ast }$, derived from the joint
velocities $v$ by multiplying them with corresponding moments of
inertia.

\subsection{Configuration Manifold of Realistic Human Motion}

\begin{figure}[tbh]
\centering \includegraphics[width=12cm]{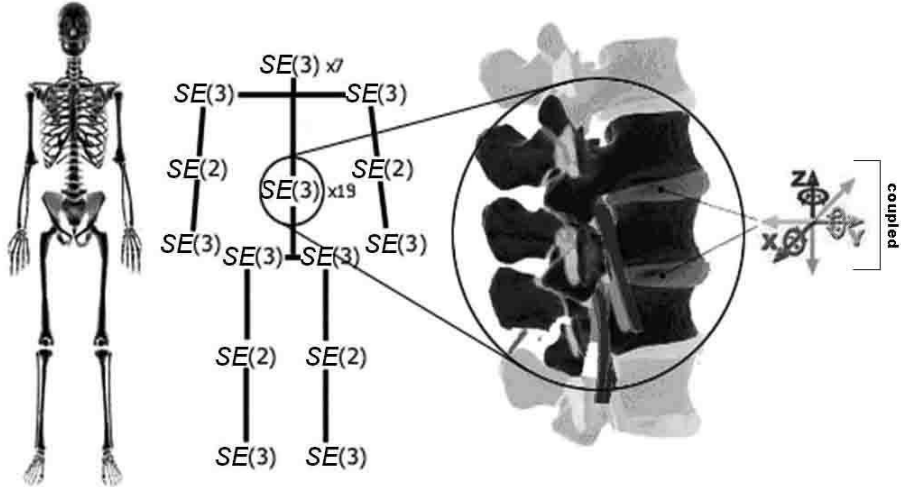} \caption{The
configuration manifold $Q_{hum}$ of the human body is defined as a
topological product of constrained $SE(3)$ groups acting in all
major (synovial) human joints, $Q_{hum}=\prod_{i}SE(3)^{i}$.}
\label{SpineSE(3)}
\end{figure}

On the other hand, human joints are more flexible than robot joints. Namely,
every rotation in all synovial human joints is followed by the corresponding
micro--translation, which occurs after the rotational amplitude is reached
\cite{VladSanjeev}. So, representation of human motion is rigorously defined
in terms of {Euclidean} $SE(3)$--groups of full rigid--body motion \cite%
{Marsden,GaneshSprSml,GaneshSprBig,GaneshADG} in all main human joints (see
Figure \ref{SpineSE(3)}). Therefore, the configuration manifold $Q_{hum}$
for human dynamics is defined as a topological product of all included
constrained $SE(3)$ groups, $Q_{rob}=\prod_{i}SE(3)^{i}$. Consequently, the
natural stage for autonomous Lagrangian dynamics of human motion is the
tangent bundle $TQ_{hum}$ \cite{LieLagr}, and for the corresponding
autonomous Hamiltonian dynamics is the cotangent bundle $T^{\ast }Q_{hum}$
\cite{IJMMS1,SIAM,IJMMS2}.

Briefly, the Euclidean SE(3)--group is defined as a semidirect
(noncommutative) product of 3D rotations and 3D translations, $%
SE(3):=SO(3)\rhd \mathbb{R}^{3}$. Its most important subgroups are the
following (for technical details see Appendix, as well as \cite%
{GaneshSprBig,ParkChung,GaneshADG}):

{{\frame{$%
\begin{array}{cc}
\mathbf{Subgroup} & \mathbf{Definition} \\ \hline
\begin{array}{c}
SO(3),\text{ group of rotations} \\
\text{in 3D (a spherical joint)}%
\end{array}
&
\begin{array}{c}
\text{Set of all proper orthogonal } \\
3\times 3-\text{rotational matrices}%
\end{array}
\\ \hline
\begin{array}{c}
SE(2),\text{ special Euclidean group} \\
\text{in 2D (all planar motions)}%
\end{array}
&
\begin{array}{c}
\text{Set of all }3\times 3-\text{matrices:} \\
\left[
\begin{array}{ccc}
\cos \theta & \sin \theta & r_{x} \\
-\sin \theta & \cos \theta & r_{y} \\
0 & 0 & 1%
\end{array}%
\right]%
\end{array}
\\ \hline
\begin{array}{c}
SO(2),\text{ group of rotations in 2D} \\
\text{subgroup of }SE(2)\text{--group} \\
\text{(a revolute joint)}%
\end{array}
&
\begin{array}{c}
\text{Set of all proper orthogonal } \\
2\times 2-\text{rotational matrices} \\
\text{ included in }SE(2)-\text{group}%
\end{array}
\\ \hline
\begin{array}{c}
\mathbb{R}^{3},\text{ group of translations in 3D} \\
\text{(all spatial displacements)}%
\end{array}
& \text{Euclidean 3D vector space}%
\end{array}%
$}}}

\subsection{The Covariant Force Law and Mechanics of Musculoskeletal Injury}

\begin{figure}[h]
\centerline{\includegraphics[width=7cm]{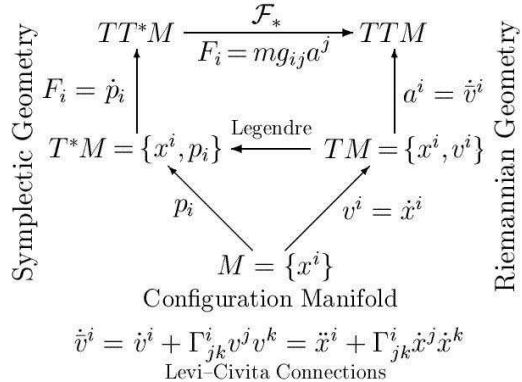}}
\caption{{\protect\small Riemannian--symplectic geometry of the covariant
force law.}}
\label{CovForce}
\end{figure}

The SE(3)--dynamics applied to human body gives the fundamental law of
biomechanics, which is the {covariant force law} \cite%
{GaneshSprSml,GaneshWSc,GaneshSprBig,GaneshADG}. It states:
\begin{equation*}
\text{Force co-vector field}=\text{Mass distribution}\times \text{%
Acceleration vector field},
\end{equation*}
which is formally written (using Einstein's summation convention over
repeating indices, with indices labelling the three Cartesian
(X-Y-Z)--translations and the corresponding three Euler angles):
\begin{equation*}
F_{{\mu}}=m_{{\mu}{\nu}}a^{{\nu}},\qquad ({\mu,\nu}=1,...,6)
\end{equation*}
where $F_{{\mu}}$ denotes the 6 covariant components of the external
``pushing''\ SE(3)--force co-vector field, $m_{{\mu}{\nu}}$ represents the 6$%
\times $6 covariant components of proximal segment's inertia--metric tensor,
while $a^{{\nu}}$ corresponds to the 6 contravariant components of the
segment's internal SE(3)--acceleration vector-field. This law states that
contrary to common perception, acceleration and force are not quantities of
the same nature: while acceleration is a non-inertial vector field, force is
an inertial co-vector field. This apparently insignificant difference
becomes crucial in injury prediction/prevention, as formalized below.
Geometrical elaboration of the covariant force law (briefly shown in Figure %
\ref{CovForce}) is fully elaborated in \cite%
{GaneshSprSml,GaneshWSc,GaneshSprBig,GaneshADG}

Now we come to injury prediction. It was shown in \cite%
{GaneshSprSml,GaneshWSc} that the general cause spinal and other
musculoskeletal injuries is the {SE(3)--jolt}, which is a sharp
and sudden change in the SE(3)--force acting on the mass--inertia
distribution of the proximal segment to the injured joint. The
SE(3)--jolt is a `delta'--change in a total 3D force--vector
acting on joint coupled to a total 3D torque--vector. In other
words, the SE(3)--jolt is a sudden, sharp and discontinues shock
in all 6 coupled DOF, distributed along the three Cartesian
($x,y,z$)--translations and the three corresponding Euler angles
around the Cartesian axes: roll, pitch and yaw. The SE(3)--jolt is
rigorously defined in terms of differential geometry \cite%
{GaneshSprBig,GaneshADG}. Briefly, it is the absolute time--derivative of
the covariant force 1--form acting on the joint.

Formally, the covariant (absolute, Bianchi) time-derivative $\frac{{D}}{dt}%
(\cdot )$ of the covariant SE(3)--force $F_{{\mu }}$ defines the
corresponding external ``striking" SE(3)--jolt co-vector field:
\begin{equation}
\frac{{D}}{dt}(F_{{\mu }})=m_{{\mu }{\nu }}\frac{{D}}{dt}(a^{{\nu }})=m_{{%
\mu }{\nu }}\left( \dot{a}^{{\nu }}+\Gamma _{\mu \lambda }^{{\nu }}a^{{\mu }%
}a^{{\lambda }}\right) ,  \label{Bianchi}
\end{equation}%
where ${\frac{{D}}{dt}}{(}a^{{\nu }})$ denotes the 6 contravariant
components of the proximal segment's internal SE(3)--jerk vector-field and
overdot ($\dot{~}$) denotes the time derivative. $\Gamma _{\mu \lambda }^{{%
\nu }}$ are the Christoffel's symbols of the Levi--Civita connection for the
SE(3)--group, which are zero in case of pure Cartesian translations and
nonzero in case of rotations as well as in the full--coupling of
translations and rotations.

\begin{figure}[h]
\centerline{\includegraphics[width=3cm]{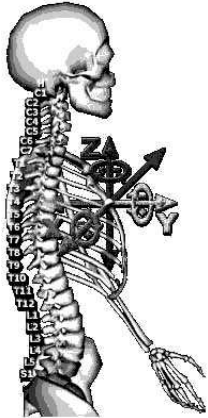}}
\caption{{\protect\small Reference frame of the Human Biodynamics Engine
(HBE). The purpose of the HBE simulator is prediction of the risk of soft
spinal and other musculo-skeletal injuries, as well as biodynamical behavior
modelling.}}
\label{HBERefFrame2}
\end{figure}

In particular, the spine, or vertebral column, dynamically represents a
chain of 26 movable vertebral bodies, joint together by transversal
viscoelastic intervertebral discs and longitudinal elastic tendons (see
Figure \ref{HBERefFrame2}). Textbooks on functional anatomy describe the
following spinal movements: (a) cervical intervertebral joints allow all
three types of movements: flexion and extension (in the sagittal plane),
lateral flexion (in the frontal plane) and rotation (in the transverse
plane); (b) thoracic joints allow rotation and lateral flexion (limited by
ribs), while flexion/extension is prevented; and (c) lumbar joints allow
flexion/extension as well as limited lateral flexion, while rotation is
prevented. This popular picture is fine for the description of safe spinal
movements; however, to be able to predict and prevent spinal injuries (both
soft ones related to the back-pain syndrome and hard ones related to discus
hernia), which are in the domain of unsafe intervertebral movements, a much
more rigorous description is needed. The main cause of spinal injuries is
the SE(3)--jolt, a shock that breaks the spinal structure and/or function.

\subsection{Lagrangian Formulation of Biodynamics}

The general form of Lagrangian human/humanoid biodynamics on the
corresponding Riemannian tangent bundles $TQ_{rob}$ and $TQ_{hum}$ of the
configuration manifolds $Q_{rob}$ and $Q_{hum}$ (precisely derived in \cite%
{LieLagr,GaneshSprSml,GaneshSprBig}) can be formulated in a unified form as:
\begin{equation}
\frac{d}{dt}L_{\dot{x}^{i}}-L_{x^{i}}=\mathcal{F}_{i}\left( t,x,\dot{x}%
\right) ,\qquad (i=1,...,n)  \label{classic}
\end{equation}%
where $n$ denotes the number of DOF for both $n_{hum}$ and $n_{rob}$, $%
L=L(t,x,\dot{x}):TQ\rightarrow \mathbb{R}$ is the human/humanoid {%
Lagrangian} function, defined on the $(2n+1)$--dimensional {jet
manifolds},\footnote{%
In mechanics, we consider a pair of maps
$f_{1},f_{2}:\mathbb{R}\rightarrow Q $ from the real line
$\mathbb{R}$, representing the {time} $t-$axis, into a smooth $n$D
{configuration manifold} $Q$. We say that the two
maps $f_{1}=f_{1}(t)$ and $f_{2}=f_{2}(t)$ have the same $k-$jet {$%
j_{t}^{k}f $} at a specified time instant $t_{0}\in \mathbb{R}$, iff:
\par
\begin{enumerate}
\item $f_{1}(t)=f_{2}(t)$ at $t_{0}\in \mathbb{R}$; and also
\par
\item the first $k$ terms of their Taylor--series expansions around $%
t_{0}\in \mathbb{R}$ are equal.
\end{enumerate}
\par
The set of all $k-$jets $j_{t}^{k}f:\mathbb{R}\rightarrow Q$ is the $k-$jet
manifold $J^{k}(\mathbb{R},Q)$. In particular, $J^{1}(\mathbb{R},Q)\cong
\mathbb{R}\times TQ$ (for technical details, see \cite%
{GaneshSprBig,GaneshADG}).} $X_{rob}=J_{rob}^{1}(\mathbb{R},Q_{rob})\cong
\mathbb{R}\times TQ_{rob}$ and $X_{hum}=J_{hum}^{1}(\mathbb{R},Q_{hum})\cong
\mathbb{R}\times TQ_{hum}$, respectively, with local canonical variables $%
(t;x_{rob}^{i};\dot{x}_{rob}^{i})$ and $(t;x_{hum}^{i};\dot{x}_{hum}^{i})$,
respectively. Its coordinate and velocity partial derivatives are
respectively denoted by $L_{x^{i}}$ and $L_{\dot{x}^{i}}$.

\subsection{Local Muscular Mechanics}

The right--hand side terms $\mathcal{F}_{i}(t,x,\dot{x})$ of
(\ref{classic}) denote any type of {external} torques and forces,
including excitation and contraction dynamics of
muscular--actuators and rotational dynamics of hybrid robot
actuators, as well as (nonlinear) dissipative joint torques and
forces and external stochastic perturbation torques and forces. In
particular, we have
\cite{LieLagr,GaneshSprSml,GaneshWSc,GaneshSprBig}):

{1. Synovial joint dynamics}, giving the first stabilizing effect
to the conservative skeleton dynamics, is described by the
$(x,\dot{x})$--form of the {Rayleigh -- Van der Pol's dissipation
function}
\begin{equation*}
R=\frac{1}{2}\sum_{i=1}^{n}\,(\dot{x}^{i})^{2}\,[\alpha _{i}\,+\,\beta
_{i}(x^{i})^{2}],\quad
\end{equation*}
where $\alpha _{i}$ and $\beta _{i}$ denote dissipation parameters. Its
partial derivatives give rise to the viscous--damping torques and forces in
the joints
\begin{equation*}
\mathcal{F}_{i}^{joint}=\partial R/\partial \dot{x}^{i},
\end{equation*}
which are linear in $\dot{x}^{i}$ and quadratic in $x^{i}$.

{2. Muscular dynamics}, giving the driving torques and forces $%
\mathcal{F}_{i}^{muscle}=\mathcal{F}_{i}^{muscle}(t,x,\dot{ x})$ with $%
(i=1,\dots ,n)$ for RHB, describes the internal {excitation} and
{contraction} dynamics of {equivalent muscular actuators} \cite%
{Hatze}.

(a) {Excitation dynamics} can be described by an impulse {%
force--time} relation
\begin{eqnarray*}
F_{i}^{imp} &=&F_{i}^{0}(1\,-\,e^{-t/\tau _{i}})\text{ \qquad if stimulation
}>0 \\
\quad F_{i}^{imp} &=&F_{i}^{0}e^{-t/\tau _{i}}\qquad \qquad \quad\text{if
stimulation }=0,\quad
\end{eqnarray*}
where $F_{i}^{0}$ denote the maximal isometric muscular torques
and forces, while $\tau _{i}$ denote the associated time
characteristics of particular muscular actuators. This relation
represents a solution of the Wilkie's muscular {active--state
element} equation \cite{Wilkie}
\begin{equation*}
\dot{\mu}\,+\,\gamma \,\mu \,=\,\gamma \,S\,A,\quad \mu (0)\,=\,0,\quad
0<S<1,
\end{equation*}
where $\mu =\mu (t)$ represents the active state of the muscle, $\gamma $
denotes the element gain, $A$ corresponds to the maximum tension the element
can develop, and $S=S(r)$ is the `desired' active state as a function of the
motor unit stimulus rate $r$. This is the basis for the RHB force controller.

(b) {Contraction dynamics} has classically been described by the
Hill's {hyperbolic force--velocity }relation \cite{Hill}
\begin{equation*}
F_{i}^{Hill}\,=\,\frac{\left( F_{i}^{0}b_{i}\,-\,\delta _{ij}a_{i}\dot{x}%
^{j}\,\right) }{\left( \delta _{ij}\dot{x}^{j}\,+\,b_{i}\right) },\,\quad
\end{equation*}
where $a_{i}$ and $b_{i}$ denote the {Hill's parameters},
corresponding to the energy dissipated during the contraction and
the phosphagenic energy conversion rate, respectively, while
$\delta _{ij}$ is the Kronecker's $\delta-$tensor.

In this way, RHB describes the excitation/contraction dynamics for the $i$th
equivalent muscle--joint actuator, using the simple impulse--hyperbolic
product relation
\begin{equation*}
\mathcal{F}_{i}^{muscle}(t,x,\dot{x})=\,F_{i}^{imp}\times F_{i}^{Hill}.\quad
\end{equation*}

Now, for the purpose of biomedical engineering and rehabilitation,
RHB has developed the so--called {hybrid rotational actuator}. It
includes, along with muscular and viscous forces, the D.C. motor
drives, as used in robotics \cite{Vuk,LieLagr,GaneshSprSml}
\begin{equation*}
\mathcal{F}_{k}^{robo}=i_{k}(t)-J_{k}\ddot{x}_{k}(t)-B_{k}\dot{x}_{k}(t),
\end{equation*}
with
\begin{equation*}
l_{k}i_{k}(t)+R_{k}i_{k}(t)+C_{k}\dot{x}_{k}(t)=u_{k}(t),
\end{equation*}
where $k=1,\dots,n$, $i_{k}(t)$ and $u_{k}(t)$ denote currents and voltages
in the rotors of the drives, $R_{k},l_{k}$ and $C_{k}$ are resistances,
inductances and capacitances in the rotors, respectively, while $J_{k}$ and $%
B_{k}$ correspond to inertia moments and viscous dampings of the drives,
respectively.

Finally, to make the model more realistic, we need to add some stochastic
torques and forces \cite{GaneshIEEE,NeuFuz}
\begin{equation*}
\mathcal{F}_{i}^{stoch}=B_{ij}[x^{i}(t),t]\,dW^{j}(t)
\end{equation*}
where $B_{ij}[x(t),t]$ represents continuous stochastic {diffusion
fluctuations}, and $W^{j}(t)$ is an $N-$variable {Wiener process}
(i.e. generalized Brownian motion), with
$dW^{j}(t)=W^{j}(t+dt)-W^{j}(t)$ for $j=1,\dots,N$.

\subsection{Hamiltonian Biodynamics and Reflex Servo--Control}

The general form of Hamiltonian human/humanoid biodynamics on the
corresponding symplectic cotangent bundles $T^{\ast }Q_{rob}$ and
$T^{\ast }Q_{hum}$ of the configuration manifolds $Q_{rob}$ and
$Q_{hum}$ (derived in \cite{IJMMS2,VladSanjeev,GaneshSprSml}) is
based on the {affine Hamiltonian function} $H_{a}:T^{\ast
}Q\rightarrow \mathbb{R},$ in local canonical coordinates on
$T^{\ast }Q$ given as
\begin{equation}
H_{a}(x,p,u)=H_{0}(x,p)-H^{j}(x,p)\,u_{j},  \label{aff}
\end{equation}%
where $H_{0}(x,p)$ is the physical Hamiltonian (kinetic + potential energy)
dependent on joint coordinates $x^{i}$ and canonical momenta $p^{i}$, $%
H^{j}=H^{j}(x,p)$, ($j=1,\dots ,\,m\leq n$ are the coupling Hamiltonians
corresponding to the system's active joints and $u_{i}=u_{i}(t,x,p)$ are
(reflex) feedback--controls. Using (\ref{aff}) we come to the affine
Hamiltonian control HBE--system, in deterministic form%
\begin{align}
\dot{x}^{i}& =\partial _{p_{i}}H_{0}-\partial _{p_{i}}H^{j}\,u_{j}+\partial
_{p_{i}}R,  \label{af1} \\
\dot{p}_{i}& =\mathcal{F}_{i}-\partial _{x^{i}}H_{0}+\partial
_{x^{i}}H^{j}\,u_{j}+\partial _{x^{i}}R,  \notag \\
o^{i}& =-\partial _{u_{i}}H_{a}=H^{j},  \notag \\
x^{i}(0)& =x_{0}^{i},\qquad p_{i}(0)=p_{i}^{0},  \notag \\
(i& =1,\dots ,n;\qquad j=1,\dots ,\,Q\leq n),  \notag
\end{align}%
(where $\partial _{u}\equiv \partial /\partial u$, $\mathcal{F}_{i}=\mathcal{%
F}_{i}(t,x,p),$ $H_{0}=H_{0}(x,p),$ $H^{j}=H^{j}(x,p),$ $H_{a}=H_{a}(x,p,u),$
$R=R(x,p)$), as well as in the fuzzy--stochastic form \cite%
{GaneshIEEE,NeuFuz}%
\begin{align}
dq^{i}& =\left( \partial _{p_{i}}H_{0}(\sigma _{\mu })-\partial
_{p_{i}}H^{j}(\sigma _{\mu })\,u_{j}+\partial _{p_{i}}R\right) \,dt,  \notag
\\
dp_{i}& =B_{ij}[x^{i}(t),t]\,dW^{j}(t)\qquad +\qquad \qquad   \label{af2} \\
& \left( \bar{\mathcal{F}}_{i}-\partial _{x^{i}}H_{0}(\sigma _{\mu
})+\partial _{x^{i}}H^{j}(\sigma _{\mu })\,u_{j}+\partial _{x^{i}}R\right)
\,dt,  \notag \\
d\bar{o}^{i}& =-\partial _{u_{i}}H_{a}(\sigma _{\mu })\,dt=H^{j}(\sigma
_{\mu })\,dt,\qquad \qquad   \notag \\
x^{i}(0)& =\bar{x}_{0}^{i},\qquad p_{i}(0)=\bar{p}_{i}^{0}\qquad \qquad
\notag
\end{align}%
In (\ref{af1})--(\ref{af2}), $R=R(x,p)$ denotes the joint (nonlinear)
dissipation function, $o^{i}$ are affine system outputs (which can be
different from joint coordinates); $\{\sigma \}_{\mu }$ \ (with $\mu \geq 1$%
) denote fuzzy sets of conservative parameters (segment lengths, masses and
moments of inertia), dissipative joint dampings and actuator parameters
(amplitudes and frequencies), while the bar $\bar{(.)}$ over a variable
denotes the corresponding fuzzified variable; $B_{ij}[q^{i}(t),t]$ denote
diffusion fluctuations and $W^{j}(t)$ are discontinuous jumps as the $n$%
--dimensional Wiener process.

In this way, the {force HBE servo--controller} is formulated as
affine control Hamiltonian--systems (\ref{af1}--\ref{af2}), which
resemble an {autogenetic motor servo} \cite{Houk}, acting on the
spinal--reflex level of the human locomotion control. A voluntary
contraction force $F$ of human skeletal muscle is reflexly excited
(positive feedback $+F^{-1}$) by the responses of its {spindle
receptors} to stretch and is reflexly inhibited (negative feedback
$-F^{-1}$) by the responses of its {Golgi tendon organs} to
contraction. Stretch and unloading reflexes are mediated by
combined actions of several autogenetic
neural pathways, forming the so--called $\mathbf{`}$motor servo.' The term $%
\mathbf{`}$autogenetic' means that the stimulus excites receptors located in
the same muscle that is the target of the reflex response. The most
important of these muscle receptors are the primary and secondary endings in
the muscle--spindles, which are sensitive to length change -- positive
length feedback $+F^{-1}$, and the Golgi tendon organs, which are sensitive
to contractile force -- negative force feedback $-F^{-1}$.

The gain $G$ of the length feedback $+F^{-1}$ can be expressed as
the {positional stiffness} (the ratio $G\approx S=dF/dx$ of the
force--$F$ change to the length--$x$ change) of the muscle system.
The greater the stiffness $S$, the less the muscle will be
disturbed by a change in load. The autogenetic circuits $+F^{-1}$
and $-F^{-1}$ appear to function as {servoregulatory loops} that
convey continuously graded amounts of excitation and inhibition to
the large ({alpha}) skeletomotor neurons. Small ({gamma})
fusimotor neurons innervate the contractile poles of muscle
spindles and function to modulate spindle--receptor discharge.

\subsection{Cerebellum--Like Velocity and Jerk Control}

{Nonlinear velocity} and {jerk} (time derivative of acceleration)
{servo--controllers} \cite{StrAttr}, developed in HBE using the
Lie--derivative formalism, resemble self--stabilizing and adaptive
tracking action of the cerebellum \cite{HoukBarto}. By introducing
the vector--fields $f$ and $g$, given respectively by
\begin{equation*}
f=\left( \partial _{p_{i}}H_{0},\,-\partial _{q^{i}}H_{0}\right) ,\qquad
g=\left( -\partial _{p_{i}}H^{j},\,\partial _{q^{i}}H^{j}\right) ,
\end{equation*}%
we obtain the affine controller in the standard nonlinear MIMO--system form
(see \cite{Isidori,Schaft,GaneshSprBig})
\begin{equation}
\dot{x}_{i}=f(x)+g(x)\,u_{j}.  \label{MIMO}
\end{equation}

Finally, using the {Lie derivative formalism} \cite{GaneshADG,ComDyn}%
\footnote{%
Let $F(M)$ denote the set of all smooth (i.e., $C^{\infty }$) real
valued functions $f:M\rightarrow \mathbb{R}$ on a smooth manifold
$M$, $V(M)$ -- the set of all smooth vector--fields on $M$, and
$V^{\ast }(M)$ -- the set of all differential one--forms on $M$.
Also, let the vector--field $\zeta \in V(M)$ be given with its
local flow $\phi _{t}:M\rightarrow M$ such that at a point $x\in
M$, $\frac{d}{dt}|_{t=0}\,\phi _{t}x=\zeta (x)$, and $\phi
_{t}^{\ast }$ representing the {pull--back} by $\phi _{t}$. The
{Lie derivative} differential operator $\mathcal{L}_{\zeta }$ is
defined:
\par
(i) on a function $f\in F(M)$ as
\begin{equation*}
\mathcal{L}_{\zeta }:F(M)\rightarrow F(M),\qquad \mathcal{L}_{\zeta }f=\frac{%
d}{dt}(\phi _{t}^{\ast }f)|_{t=0},
\end{equation*}%
\par
(ii) on a vector--field $\eta \in V(M)$ as
\begin{equation*}
\mathcal{L}_{\zeta }:V(M)\rightarrow V(M),\qquad \mathcal{L}_{\zeta }\eta =%
\frac{d}{dt}(\phi _{t}^{\ast }\eta )|_{t=0}\equiv \lbrack \zeta ,\eta ]
\end{equation*}%
-- the {Lie bracket}, and
\par
(iii) on a one--form $\alpha \in V^{\ast }(M)$ as
\begin{equation*}
\mathcal{L}_{\zeta }:V^{\ast }(M)\rightarrow V^{\ast }(M),\qquad \mathcal{L}%
_{\zeta }\alpha =\frac{d}{dt}(\phi _{t}^{\ast }\alpha )|_{t=0}.
\end{equation*}%
In general, for any smooth tensor field $\mathbf{T}$ on $M$, the
Lie derivative $\mathcal{L}_{\zeta }\mathbf{T}$ geometrically
represents a directional derivative of $\mathbf{T}$ along the flow
$\phi_{t}$.} and applying the {constant relative degree} $r$ to
all HB joints, the
{control law} for asymptotic tracking of the reference outputs $%
o_{R}^{j}=o_{R}^{j}(t)$ could be formulated as (generalized from \cite%
{Isidori})
\begin{equation}
u_{j}=\frac{\dot{o}_{R}^{(r)j}-L_{f}^{(r)}H^{j}+%
\sum_{s=1}^{r}c_{s-1}(o_{R}^{(s-1)j}-L_{f}^{(s-1)}H^{j})}{%
L_{g}L_{f}^{(r-1)}H^{j}},  \label{CtrLaw}
\end{equation}
where $c_{s-1}$ are the coefficients of the linear differential
equation of order $r$ for the {error function}
$e(t)=x^{j}(t)-o_{R}^{j}(t)$
\begin{equation*}
e^{(r)}+c_{r-1}e^{(r-1)}+\dots+c_{1}e^{(1)}+c_{0}e=0.
\end{equation*}

The affine nonlinear MIMO control system (\ref{MIMO}) with the
Lie--derivative control law (\ref{CtrLaw}) resembles the self--stabilizing
and synergistic output tracking action of the human cerebellum \cite%
{NeuFuz,ComDyn}. To make it adaptive (and thus more realistic),
instead of the `rigid' controller (\ref{CtrLaw}), we can use the
{adaptive Lie--derivative controller}, as explained in the seminal
paper on geometrical nonlinear control \cite{SI}.

\subsection{Cortical--Like Fuzzy--Topological Control}

For the purpose of our cortical control, the dominant, rotational part of
the human configuration manifold $M^{N}$, could be first, reduced to an $N$--%
{torus}, and second, transformed to an $N$--{cube}
(`hyper--joystick'), using the following topological techniques (see \cite%
{GaneshSprBig,GaneshADG,ComDyn}).

Let $S^{1}$ denote the constrained unit circle in the complex
plane, which is an Abelian Lie group. Firstly, we propose two
reduction homeomorphisms, using the Cartesian product of the
constrained $SO(2)-$groups:
\begin{equation*}
SO(3)\approx SO(2)\times SO(2)\times SO(2)\qquad\text{and} \qquad
SO(2)\approx S^{1}.
\end{equation*}

Next, let $I^{N}$ be the unit cube $[0,1]^{N}$ in $\mathbb{R}^{N}$ and `$%
\sim $' an equivalence relation on $\mathbb{R}^{N}$ obtained by `gluing'
together the opposite sides of $I^{N}$, preserving their orientation.
Therefore, $M^{N}$ can be represented as the quotient space of $\mathbb{R}%
^{N}$ by the space of the integral lattice points in $\mathbb{R}^{N}$, that
is an oriented and constrained $N$--dimensional torus $T^{N}$:
\begin{equation}
{\mathbb{R}^{N}/{Z}^{N}}\approx \,\prod_{i=1}^{N}S_{i}^{1}\equiv
\{(q^{i},\,i=1,\dots ,N):\mbox{mod}2\pi \}=T^{N}.  \label{torus}
\end{equation}%
Its {Euler--Poincar\'{e} characteristic} is (by the {De Rham
theorem}) both for the configuration manifold $T^{N}$ and its {%
momentum phase--space} $T^{\ast }T^{N}$ given by (see \cite{GaneshADG})
\begin{equation*}
\chi (T^{N},T^{\ast }T^{N})=\sum_{p=1}^{N}(-1)^{p}b_{p}\,,
\end{equation*}%
where $b_{p}$ are the {Betti numbers} defined as
\begin{align*}
b^{0}& =1,\, \\
b^{1}& =N,\dots b^{p}={\binom{N}{p}},\dots b^{N-1}=N, \\
b^{N}& =1,\qquad \qquad (0\leq p\leq N).
\end{align*}

Conversely by `ungluing' the configuration space we obtain the primary unit
cube. Let `$\sim^{\ast}$' denote an equivalent decomposition or `ungluing'
relation. According to Tychonoff's {product--topology theorem} \cite%
{GaneshSprBig,GaneshADG}, for every such quotient space there exists a
`selector' such that their quotient models are homeomorphic, that is, $%
T^{N}/\sim^{\ast}\approx A^{N}/\sim^{\ast}$. Therefore $I_{q}^{N}$
represents a `selector' for the configuration torus $T^{N}$ and
can be used as an $N$--directional `$\hat{q}$--command--space' for
the {feedback
control} (FC). Any subset of degrees of freedom on the configuration torus $%
T^{N}$ representing the joints included in HB has its simple, rectangular
image in the rectified $\hat{q}$--command space -- selector $I_{q}^{N}$, and
any joint angle $q^{i}$ has its rectified image $\hat{q}^{i}$.

In the case of an end--effector, $\hat{q}^{i}$ reduces to the position
vector in external--Cartesian coordinates $z^{r}\,(r=1,\dots ,3)$. If
orientation of the end--effector can be neglected, this gives a topological
solution to the standard inverse kinematics problem.

Analogously, all momenta $\hat{p}_{i}$ have their images as rectified
momenta $\hat{p}_{i}$ in the $\hat{p}$--command space -- selector $I_{p}^{N}$%
. Therefore, the total momentum phase--space manifold $T^{\ast }T^{N}$
obtains its `cortical image' as the $\widehat{(q,p)}$--command space, a
trivial $2N$--dimensional bundle $I_{q}^{N}\times I_{p}^{N}$.

Now, the simplest way to perform the feedback FC on the cortical $\widehat{%
(q,p)}$--command space $I_{q}^{N}\times I_{p}^{N}$, and also to mimic the
cortical--like behavior, is to use the $2N$-- dimensional fuzzy--logic
controller, in much the same way as in the popular `inverted pendulum'
examples (see \cite{Kosko}).

We propose the fuzzy feedback--control map $\Xi $ that maps all the
rectified joint angles and momenta into the feedback--control one--forms
\begin{equation}
\Xi :(\hat{q}^{i}(t),\,\hat{p}_{i}(t))\mapsto u_{i}(t,q,p),  \label{map}
\end{equation}%
so that their corresponding universes of discourse, $\hat{Q}^{i}=(\hat{q}%
_{max}^{i}-\hat{q}_{min}^{i})$, $\hat{P}_{i}=(\hat{p}_{i}^{max}-\hat{p}%
_{i}^{min})$ and $\hat{U}{}_{i}=(u_{i}^{max}-u_{i}^{min})$, respectively,
are mapped as
\begin{equation}
\Xi :\prod_{i=1}^{N}\hat{Q}^{i}\times \prod_{i=1}^{N}\hat{P}_{i}\rightarrow
\prod_{i=1}^{N}{}\hat{U}{}_{i}.  \label{map1}
\end{equation}

The $2N$--dimensional map $\Xi$ (\ref{map},\ref{map1}) represents a {%
fuzzy inference system}, defined by (adapted from \cite{Taca1}):

\begin{enumerate}
\item {Fuzzification} of the crisp {rectified} and {%
discretized} angles, momenta and controls using Gaussian--bell membership
functions
\begin{equation*}
\mu _{k}(\chi )=exp[-\frac{(\chi -m_{k})^{2}}{2\sigma _{k}}],\qquad
(k=1,2,\dots ,9),
\end{equation*}%
where $\chi \in D$ is the common symbol for $\hat{q}^{i}$, $\hat{p}_{i}$ and
$u_{i}(q,p)$ and $D$ is the common symbol for $\hat{Q}^{i},\hat{P}_{i}$ and $%
{}_{i}$; the mean values $m_{k}$ of the nine partitions of each universe of
discourse $D$ are defined as $m_{k}=\lambda _{k}D+\chi _{min}$, with
partition coefficients $\lambda _{k}$ uniformly spanning the range of $D$,
corresponding to the set of nine linguistic variables $L=%
\{NL,NB,NM,NS,ZE,PS,PM$, $PB,PL\}$; standard deviations are kept constant $%
\sigma _{k}=D/9$. Using the linguistic vector $L$, the $9\times 9$ FAM
(fuzzy associative memory) matrix (a `linguistic phase--plane'), is
heuristically defined for each human joint, in a symmetrical weighted form
\begin{equation*}
\mu _{kl}=\varpi _{kl}\,exp\{-50[\lambda _{k}+u(q,p)]^{2}\},\qquad
(k,l=1,...,9)
\end{equation*}%
with weights $\varpi _{kl}\in \{0.6,0.6,0.7,0.7,0.8,0.8,0.9,0.9,1.0\}$.

\item {Mamdani inference} is used on each FAM--matrix $\mu_{kl}$
for all human joints:\newline (i) $\mu(\hat{q}^{i})$ and
$\mu(\hat{p}_{i})$ are combined inside the fuzzy IF--THEN rules
using AND (Intersection, or Minimum) operator,
\begin{equation*}
\mu_{k}[\bar{u}_{i}(q,p)]=\min_{l}\{\mu_{kl}(\hat{q}^{i}),\,\mu_{kl}(\hat {p}%
_{i})\}.
\end{equation*}
(ii) the output sets from different IF--THEN rules are then combined using
OR (Union, or Maximum) operator, to get the final output, fuzzy--covariant
torques,
\begin{equation*}
\mu[u_{i}(q,p)]=\max_{k}\{\mu_{k}[\bar{u}_{i}(q,p)]\}.
\end{equation*}

\item {Defuzzification} of the fuzzy controls $\mu \lbrack
u_{i}(q,p)] $ with the `center of gravity' method
\begin{equation*}
u_{i}(q,p)=\frac{\int \mu \lbrack u_{i}(q,p)]\,du_{i}}{\int du_{i}},
\end{equation*}%
to update the crisp feedback--control one--forms $u_{i}=u_{i}(t,q,p)$.
\end{enumerate}

Now, it is easy to make this top--level controller {adaptive},
simply by {weighting} both the above fuzzy--rules and membership
functions, by
the use of any standard competitive neural--network (see, e.g., \cite{Kosko}%
). Operationally, the construction of the cortical $\widehat{(q,p)}$%
--command space $I_{q}^{N}\times I_{p}^{N}$ and the
$2N$--dimensional feedback map $\Xi $ (\ref{map},\ref{map1}),
mimic the regulation of the {motor conditioned reflexes} by the
motor cortex \cite{HoukBarto}.

\section{HBE Simulation Examples}

\begin{figure}[h]
\centerline{\includegraphics[width=13cm]{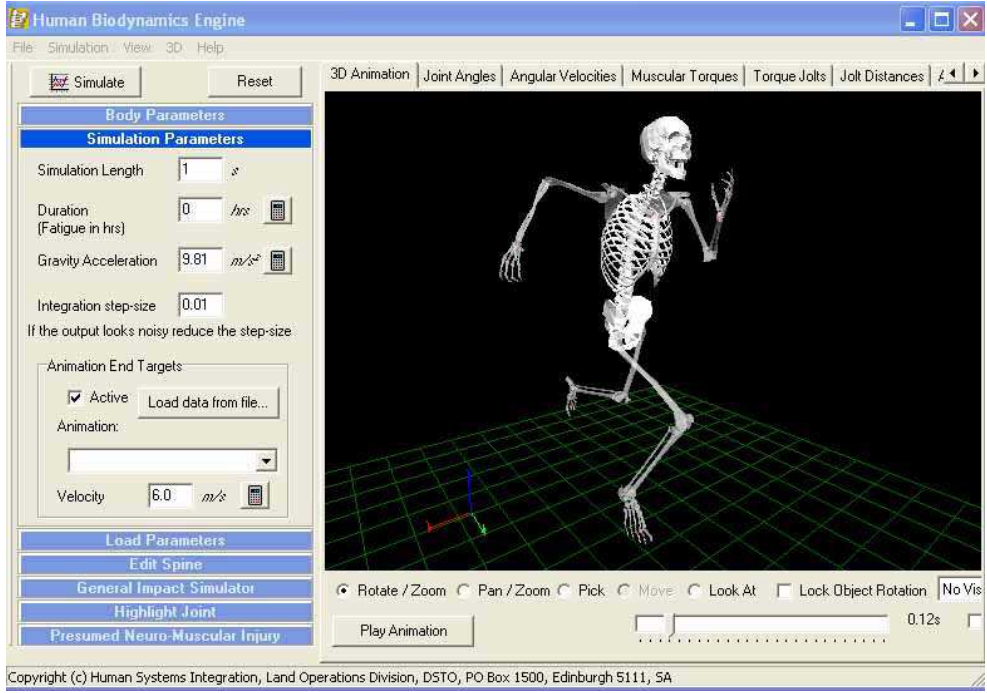}}
\caption{{\protect\small Sample output from the Human Biodynamics Engine:
running simulation with the speed of 6 m/s -- 3D animation view--port.}}
\label{RunAnim}
\end{figure}
\begin{figure}[h]
\centerline{\includegraphics[width=8cm]{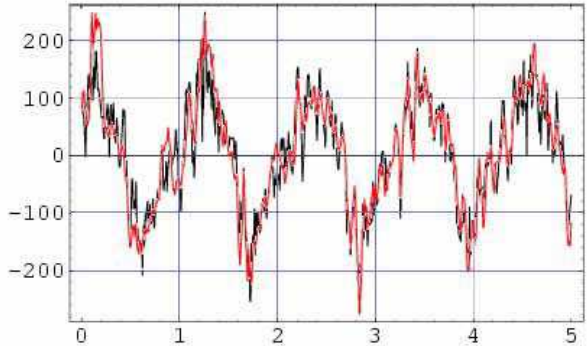}}
\caption{{\protect\small Matching the `Vicon' output with the `HBE' output
for the right-hip angular velocity around the dominant X-axis, while walking
on a treadmill with a speed of 4 km/h.}}
\label{WalkingValid}
\end{figure}
\begin{figure}[htb]
\centering \includegraphics[width=10cm]{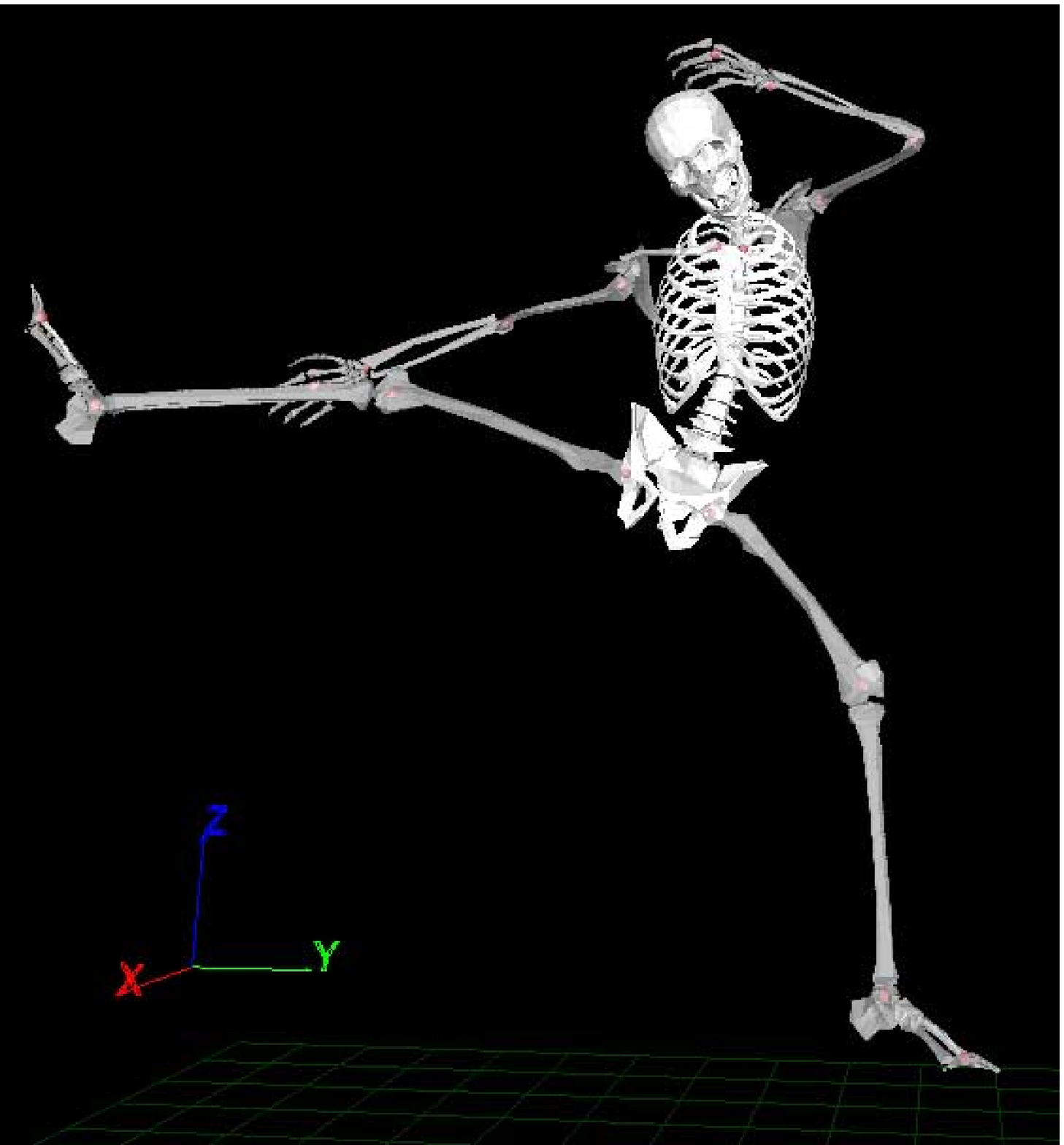}
\caption{The HBE simulating a jump-kick: a 3D viewer.}
\label{JumpKick}
\end{figure}

\begin{figure}[htb]
\centering \includegraphics[width=12cm]{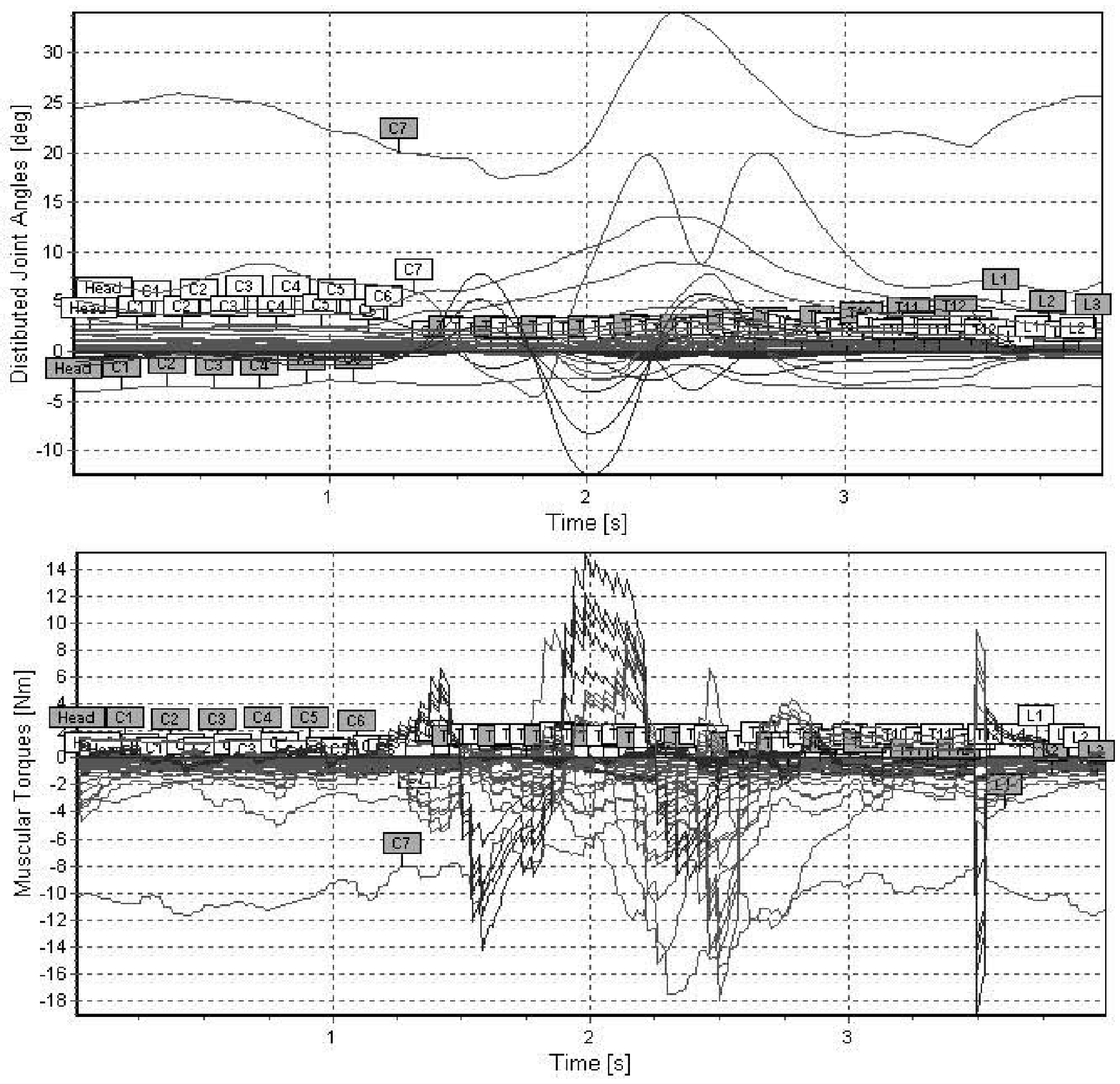}
\caption{The HBE simulating a jump-kick: calculating joint angles and
muscular torques.}
\label{AnglesTorques}
\end{figure}

The first version of the HBE simulator had the full human-like skeleton,
driven by the generalized Hamiltonian dynamics (including muscular
force-velocity and force-time curves) and two levels of reflex-like motor
control (simulated using the Lie derivative formalism) \cite%
{SIAM,GaneshSprSml,GaneshWSc}. It had 135 purely rotational DOF, strictly
following Figure \ref{SpineSO(3)}. It was created for prediction and
prevention of musculo-skeletal injuries occurring in the joints, mostly
spinal (intervertebral, see Figure \ref{HBERefFrame2}). Its performance
looked kinematically realistic, while it was not possible to validate the
driving torques. It included a small library of target movements which were
followed by the HBE's Lie--derivative controllers with efficiency of about
90\% (see Figures \ref{JumpKick} and \ref{AnglesTorques}).
\begin{figure}[htb]
\centering \includegraphics[width=8cm]{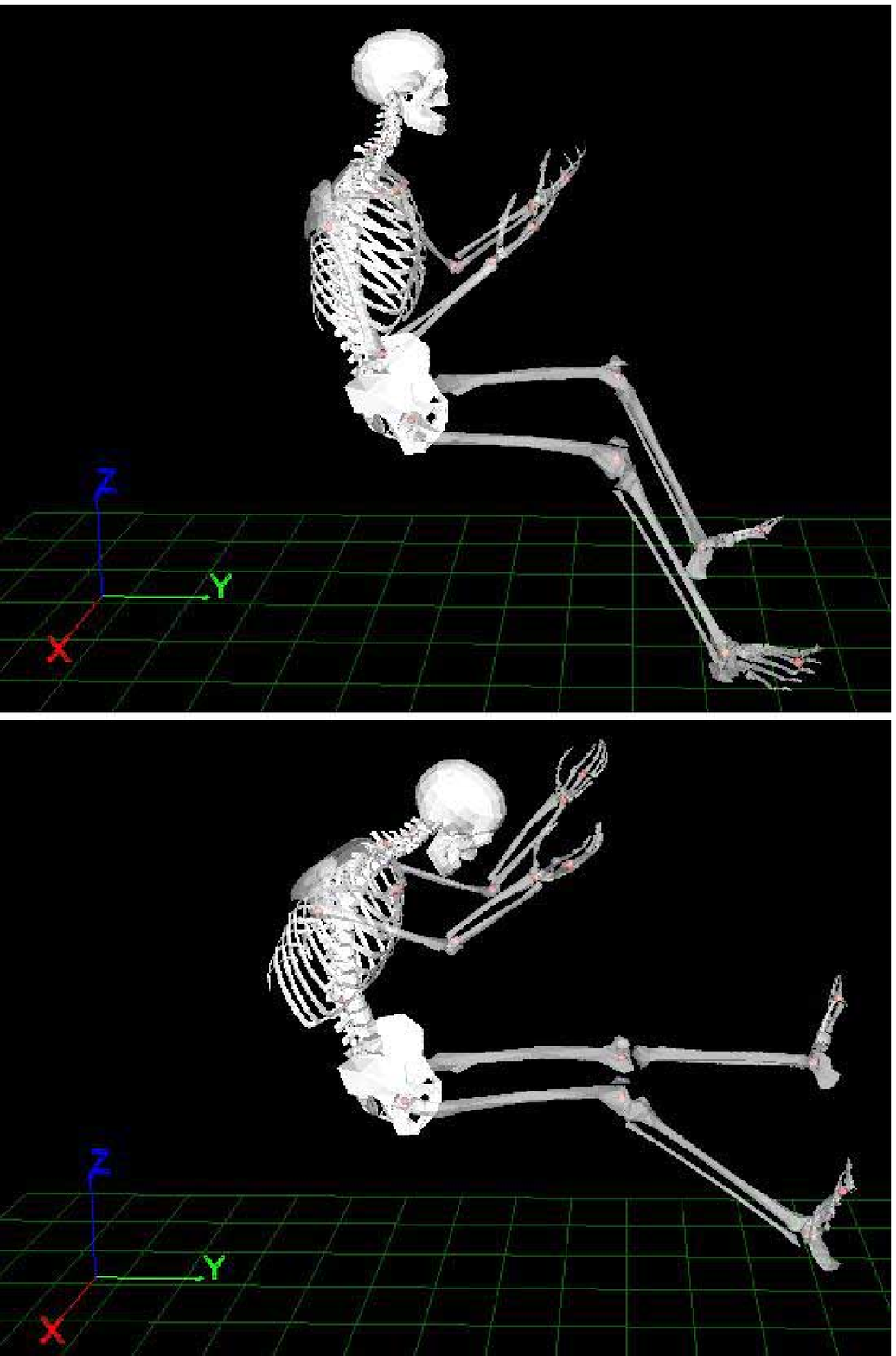}
\caption{The HBE simulating the frontal road-vehicle crash into the fixed
wall with a speed of 70 km/h: before the impact (up) and 0.12 s after the
impact.}
\label{RVcrash}
\end{figure}

\begin{figure}[htb]
\centering \includegraphics[width=12cm]{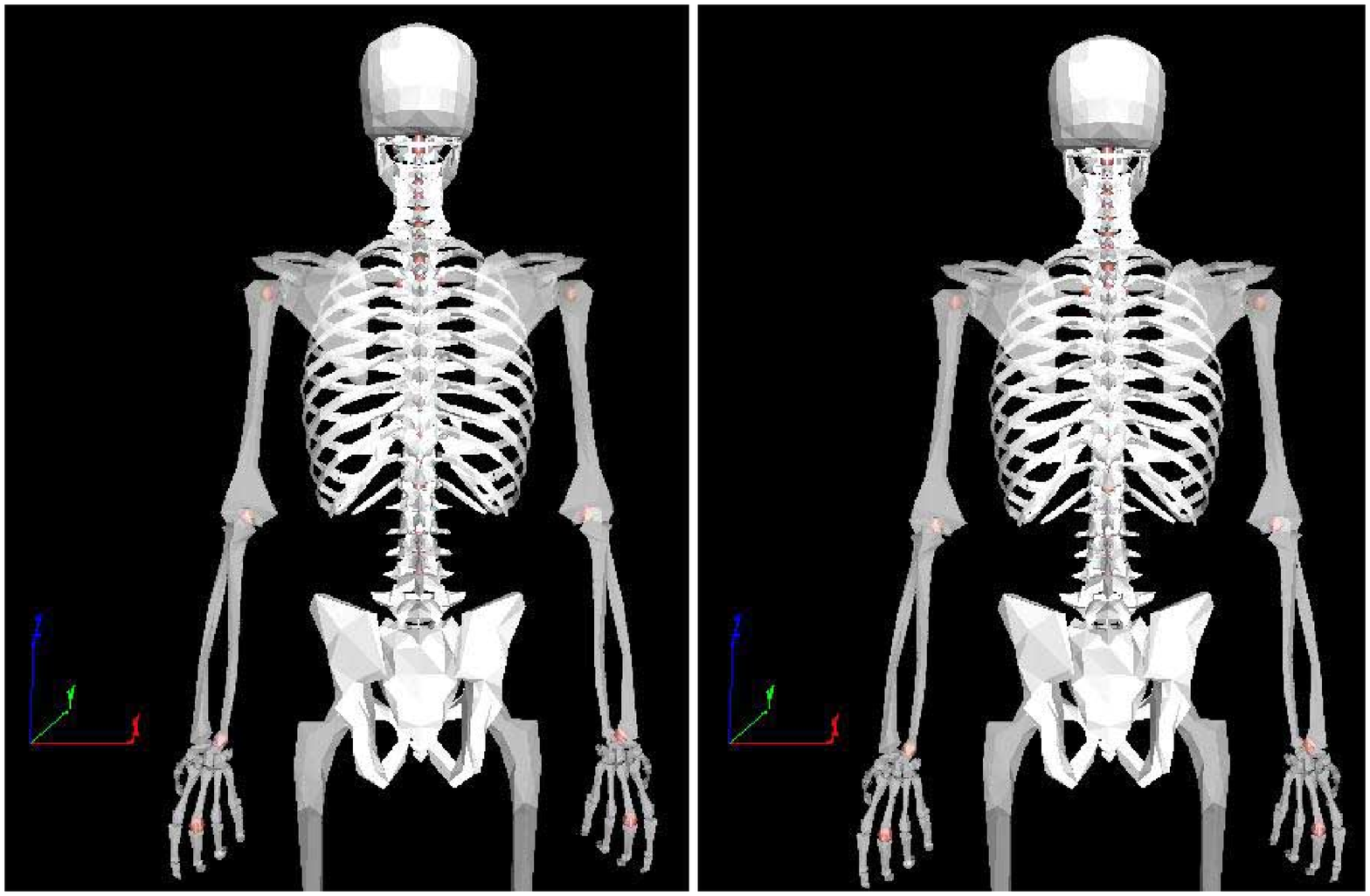}
\caption{The HBE simulating the effect of an aircraft pilot-seat ejection to
human spine compression: before the seat ejection (left) and after ejection
(right).}
\label{Ejection}
\end{figure}

\begin{figure}[htb]
\centering \includegraphics[width=12cm]{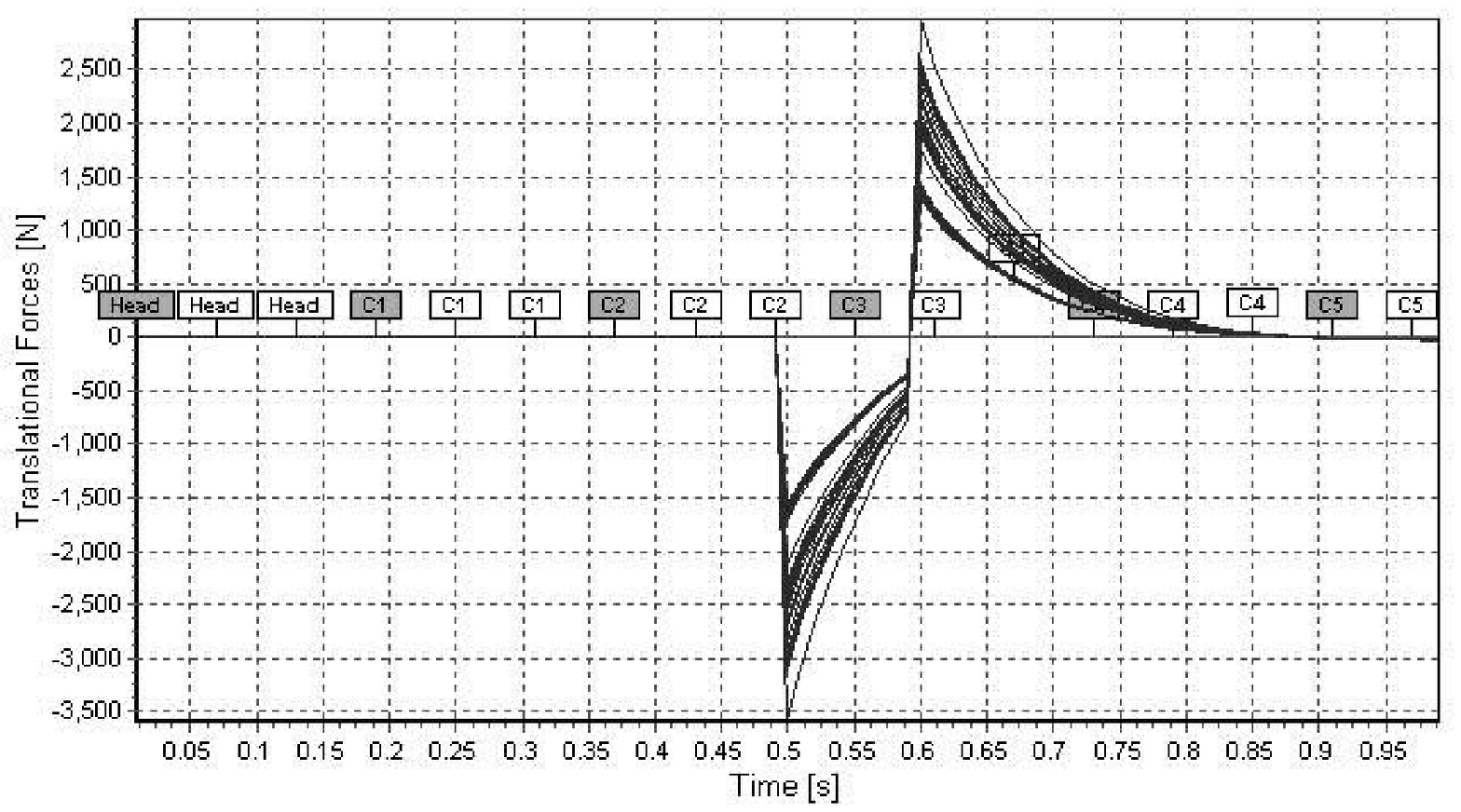}
\caption{The HBE calculating translational forces distributed along the
spinal joints during the seat ejection.}
\label{EjectForces}
\end{figure}

The HBE also includes a generic crash simulator, based on the simplified
road-vehicle impact simulator (see Figure \ref{RVcrash}). While implementing
the generic crash simulator, it became clear that purely rotational joint
dynamics would not be sufficient for the realistic prediction of
musculo-skeletal injuries. Especially, to simulate the action of a Russian
aircraft ejection-seat currently used by the American Space-shuttle, we
needed, strictly following Figure \ref{SpineSE(3)}, to implement
micro-translations in the intervertebral joints (see Figures \ref{Ejection}
and \ref{EjectForces}), as the seat provides the full body restraint and the
ejection rockets, firing with 15 g for 0.15 s -- can only compress the
spine, without any bending at all.

In this way a full rotational + translational biodynamics simulator has been
created with 270 DOF in total (not representing separate fingers). The
`HBE-simulator' has been kinematically validated \cite{HBEvalidRep} against
the standard biomechanical gait-analysis system `Vicon' \cite{Robertson}.

\section{Complexity of Biodynamical Behavior}

\subsection{Biodynamical `Entanglement'}

From the standard engineering viewpoint, having two systems combined, in the
case of biodynamics -- biological and mechanical, as a single `working
machine', we can expect that the total `machine' complexity equals the sum
of the two partial ones. For example, electrical circuitry has been a
standard modelling framework in neurophysiology (A. Hodkgin and A. Huxley
won a Nobel Prize for their circuit model of a single neuron, the celebrated
HH--neuron \cite{H-H}). Using the HH--approach for modelling human
neuro--muscular circuitry as electrical circuitry, we get an
electro--mechanical model for our bio-mechanical system, in which the
superposition of complexities is clearly valid.

On the other hand, in a recent research on dissipative quantum
brain modelling, one of the most popular issues has been {quantum
entanglement}\footnote{%
Entanglement is a term used in quantum theory to describe the way that
particles of energy/matter can become correlated to predictably interact
with each other regardless of how far apart they are; this is called a
`long--range correlation'.} between the {brain} and its {%
environment} (see \cite{MM,PV}) where the brain--environment system has an
entangled `memory' state (identified with its ground state), that cannot be
factorized into two single--mode states.\footnote{%
In the Vitiello--Pessa dissipative quantum brain model
\cite{MM,PV}, the evolution of a memory system was represented as
a trajectory of given initial condition running over
time--dependent states, each one minimizing the free energy
functional.} Similar to this microscopic brain--environment
entanglement, we conjecture the existence of a {macroscopic
neuro--mechanical entanglement} between the operating modes of our
neuro--muscular controller and purely mechanical skeleton (see
\cite{Enoka}).

In other words, we suggest that the {diffeomorphism} between the
{brain motion manifold} ($N-$cube) and the {body motion manifold}
$M^{N}$ (which can be reduced to the constrained $N-$torus),
described as the {cortical motion control}, can be considered a
`long--range correlation', thus manifesting the `biodynamical
entanglement'.

\subsection{Biodynamical Self--Assembly}

In the framework of human motion dynamics, {self--assembly}
represents {adaptive motor control}, i.e., physiological motor
training performed by {iteration of conditioned reflexes}. For
this, a combination of {supervised} and {reinforcement training}
is commonly used, in which a number of (nonlinear) {control
parameters are iteratively adjusted} similar to the weights in
neural networks, using either backpropagation--type or
Hebbian--type learning, respectively (see, e.g., \cite{Kosko}).
Every human motor skill is mastered using this general method.
Once it is mastered, it becomes {smooth and energy--efficient}, in
accordance with {Bernstein's motor coordination and dexterity}
(see \cite{Bernstein1,Bernstein1}).

Therefore, biodynamical self--assembly clearly represents an `evolution' in
the parameter--space of human motion control. One might argue that such an
evolution can be modelled using CA. However, this parameter--space, though
being a dynamical and possibly even a contractible structure, is not an
independent set of parameters -- it is necessarily coupled to the mechanical
skeleton configuration space, the plant to be controlled.

The system of 200 bones and 600 muscles can an produce infinite
number of different movements. In other words, the {output--space
dimension} of a skilled human motion dynamics equals {infinity} --
there is no upper limit to the number of possible different human
movements, starting with simple walk, run, jump, throw, play, etc.
Even for the simplest motions, like walking, a child needs about
12 months to master it (and Honda robots took a decade to achieve
this).

Furthermore, as human motion represents a simplest and yet well--defined
example of a general human behavior, it is possible that other human
behavioral and performance skills are mastered (i.e., self--assembled) in a
similar way.

\subsection{Biodynamical Synchronization}

The route to simplicity in biodynamics is {synchronization}. Both
synchronization and {phase--locking} are ubiquitous in nature as
well
as in human brain (see \cite{Izhikevich1,Izhikevich2,Izhikevich3,Izhikevich4}%
). Synchronization can occur in {cyclic forms of human motion}
(e.g., walking, running, cycling, swimming), both externally, in
the form of
{oscillatory dynamics}, and internally, in the form of {%
oscillatory cortical--control}. This oscillatory synchronization, e.g., in
walking dynamics, has three possible forms: in--phase, anti--phase, and
out--of--phase. The underlying phase--locking properties determined by type
of oscillator (e.g., periodic/chaotic, relaxation, bursting\footnote{%
Periodic bursting behavior in neurons is a recurrent transition between a
quiescent state and a state of repetitive firing. Three main types of neural
bursters are: (i) parabolic bursting (`circle/circle'), (ii) square--wave
bursting (`fold/homoclinic'), and (iii) elliptic bursting (`subHopf/fold
cycle'). Most burster models can be written in the singularly perturbed form:
\par
\begin{center}
$x = f(x, y),\qquad y = \mu g(x, y),$%
\end{center}
\par
where $x\in \mathbb{R}^m$ is a vector of fast variables
responsible for repetitive firing (e.g., the membrane voltage and
fast currents). The vector $y\in \mathbb{R}^k$ is a vector of slow
variables that modulates the firing (e.g., slow (in)activation
dynamics and changes in intracellular Ca$^{2+}$ concentration).
The small parameter $\mu<< 1$ is a ratio of fast/slow time scales.
The synchronization dynamics between bursters depends crucially on
their spiking frequencies, i.e., the interactions are most
effective when the presynaptic inter-spike frequency matches the
frequency of postsynaptic oscillations. The synchronization
dynamics between bursters in the cortical motion planner induces
synchronization dynamics between upper and lower limbs in
oscillatory motions.}, pulse-coupled, slowly connected, or
connections with time delay) involved in the cortical control
system (motion planner). According to Izhikevich--Hoppensteadt
work (ibid), phase--locking is prominent in the brain: it
frequently results in coherent activity of neurons and neuronal
groups, as seen in recordings of local field potentials and EEG.
In essence, the purpose of brain control of human motion is
reduction of mechanical configuration space; brain achieves this
through synchronization.

While cyclic movements indeed present a natural route to Biodynamical
synchronization, both on the dynamical and cortical--control level, the
various forms of synchronized group behavior in sport (such as synchronized
swimming, diving, acrobatics) or in military performance represent the
imperfect products of hard training. The synchronized team performance is
achievable, but the cost is a difficult long--term training and sacrifice of
one's natural characteristics.

\section{Conclusion}

We have presented various aspects of development of the Human
Biodynamics Engine. The HBE geometry is based on anthropomorphic
tree of Euclidean SE(3)--groups. Its dynamics was at first
Lagrangian and later changed to Hamiltonian (dynamically
equivalent, but superior for control). Its actuators are
`equivalent muscles', following classical Hill--Hatze muscular
mechanics. Its reflexes follow Houk's `autogenetic' stretch--Golgi
prescription. Its `cerebellum' is modelled using Lie-derivative
formalism. Its brain is fuzzy--topological. Its complexity shows
biodynamical `entanglement', self--assembly and oscillatory
synchronization. Its simulations demonstrate the necessity of
micro-translations in the human joints, which cannot exist in
robots. The main purpose for its development has been prediction
of neuro-musculo-skeletal injuries. For this purpose, the concept
of rotational (soft) and translational (hard) jolts has been
developed and implemented in HBE. The HBE--simulator is currently
under the thorough validation process. Kinematic validation has
mostly been completed, while for the validation of torques and
forces we are still lacking adequate \emph{in vivo }measurement
technology.

\section{Appendix: The $SE(3)-$Group of General Rigid Motions}

The special Euclidean group $SE(3):=SO(3)\rhd \mathbb{R}^{3}$,
(the semidirect product of the group of rotations with the
corresponding group of translations), is the Lie group consisting
of isometries of the Euclidean 3D space $\mathbb{R}^{3}$ (see
\cite{GaneshSprBig,ParkChung,GaneshADG}).

An element of $SE(3)$ is a pair $(A,a)$ where $A\in SO(3)$ and $a\in \mathbb{%
R}^{3}.$ The action of $SE(3)$ on $\mathbb{R}^{3}$ is the rotation $A$
followed by translation by the vector $a$ and has the expression
\begin{equation*}
(A,a)\cdot x=Ax+a.
\end{equation*}

The Lie algebra of the Euclidean group $SE(3)$ is $\mathfrak{se}(3)=\mathbb{R%
}^{3}\times \mathbb{R}^{3}$ with the Lie bracket
\begin{equation}
\lbrack (\xi ,u),(\eta ,v)]=(\xi \times \eta ,\xi \times v-\eta \times u).
\label{lbse3}
\end{equation}

Using homogeneous coordinates, we can represent $SE(3)$ as follows,
\begin{equation*}
SE(3)=\ \ \left\{ \left(
\begin{array}{cc}
R & p \\
0 & 1%
\end{array}
\right) \in GL(4,\mathbb{R}):R\in SO(3),\,p\in \mathbb{R}^{3}\right\} ,
\end{equation*}
with the action on $\mathbb{R}^{3}$ given by the usual matrix--vector
product when we identify $\mathbb{R}^{3}$ with the section $\mathbb{R}%
^{3}\times \{1\}\subset \mathbb{R}^{4}$. In particular, given
\begin{equation*}
g=\left(
\begin{array}{cc}
R & p \\
0 & 1%
\end{array}
\right) \in SE(3),
\end{equation*}
and $q\in \mathbb{R}^{3}$, we have
\begin{equation*}
g\cdot q=Rq+p,
\end{equation*}
or as a matrix--vector product,
\begin{equation*}
\left(
\begin{array}{cc}
R & p \\
0 & 1%
\end{array}
\right) \left(
\begin{array}{c}
q \\
1%
\end{array}
\right) =\left(
\begin{array}{c}
Rq+p \\
1%
\end{array}
\right) .
\end{equation*}

The Lie algebra of $SE(3)$, denoted $\mathfrak{se}(3)$, is given by \
\begin{equation*}
\mathfrak{se}(3)=\ \ \left\{ \left(
\begin{array}{cc}
\omega & v \\
0 & 0%
\end{array}
\right) \in M_{4}(\mathbb{R}):\omega\in \mathfrak{so}(3),\,v\in \mathbb{R}%
^{3}\right\} ,
\end{equation*}
where the attitude (or, angular velocity) matrix $\omega:\mathbb{R}%
^{3}\rightarrow \mathfrak{so}(3)$ is given by
\begin{equation*}
\omega=\left(
\begin{array}{ccc}
0 & -\omega _{z} & \omega _{y} \\
\omega _{z} & 0 & -\omega _{x} \\
-\omega _{y} & \omega _{x} & 0%
\end{array}
\right) .
\end{equation*}

The so--called {exponential map}, $\exp
:\mathfrak{se}(3)\rightarrow SE(3)$, is given by
\begin{equation*}
\exp \left(
\begin{array}{cc}
\omega & v \\
0 & 0%
\end{array}
\right) =\left(
\begin{array}{cc}
\exp (\omega) & Av \\
0 & 1%
\end{array}
\right) ,
\end{equation*}
where

\begin{equation*}
A=I+\frac{1-\cos \left\Vert \omega \right\Vert }{\left\Vert \omega
\right\Vert ^{2}}\omega+\frac{\left\Vert \omega \right\Vert -\sin \left\Vert
\omega \right\Vert }{\left\Vert \omega \right\Vert ^{3}} \omega^{2},
\end{equation*}
and $\exp (\omega)$ is given by the {Rodriguez' formula},
\begin{equation*}
\exp (\omega)=I+\frac{\sin \left\Vert \omega \right\Vert }{\left\Vert \omega
\right\Vert }\omega+\frac{1-\cos \left\Vert \omega \right\Vert }{\left\Vert
\omega \right\Vert ^{2}}\omega^{2}.
\end{equation*}

\end{document}